\documentstyle[prl,aps,floats,epsf,color]{revtex}
\begin{document}
\baselineskip=12pt
\def\be{\begin{equation}}
\def\ee{\end{equation}}
\def\bea{\begin{eqnarray}}
\def\eea{\end{eqnarray}}
\def\E{{\rm e}}
\def\bearst{\begin{eqnarray*}}
\def\eearst{\end{eqnarray*}}
\def\peleven{\parbox{11cm}}
\def\peffec{\peight{\bearst\eearst}\hfill\peleven}
\def\pspace{\peight{\bearst\eearst}\hfill}
\def\ptwelve{\parbox{12cm}}
\def\peight{\parbox{8mm}}
\twocolumn[\hsize\textwidth\columnwidth\hsize\csname@twocolumnfalse\endcsname

\title
{ Mid-Infrared Radiation as a Short-Term Earthquake Precursor }
\author
{ M. Allameh-Zadeh, A. Ansari, A. Bahraminasab, K. Kaviani, A.
Mahdavi Ardakani, \\ H. Mehr-nahad, D. Mehr-shahi, M.D. Niry, M.
Reza Rahimi Tabar, S. Tabatabai, \\ N. Taghavinia M. Vesaghi and
F. Zamani}
\address
{\it  Department of Physics, Sharif University of  Technology,\\
P.O. Box 11365-9161, Tehran, Iran. }


\maketitle


\begin{abstract}

Recently it has been found by F. Freund that the granite under
high pressure undergoes a phase transition from insulator to a
p-type semiconductor. This phase transition is a key concept to
understanding pre-earthquake phenomena. This effect accompanies
with the radiation of the granite in the mid-infrared region. we
were able to predict the recent earthquake in the south of Iran
by monitoring this radiation.

\end{abstract}
\hspace{.3in}
\newpage
]

Ordinary rocks are insulators. Rocks placed under great stress,
however, sometimes act like semiconductors. It has been found by
Freund that, before a quake, pairs of positive charges called
'defect electrons' or 'positive holes' split up and migrate to
the surface of stressed rocks. There they recombine with each
other and, in the process, release infrared radiation. Indeed
insulator turns into semiconductor when p-holes are activated.
This phase transition is responsible for the observation of
earthquake clouds, positive potential on the granites, magnetic
field anomaly, decreasing the surface electrical resistivity and
mid-Infrared radiation etc [1]. On 14th Feb. 2004 at 3:00 GMT, we
have observed a creation of ionized cloud in the place with
location 30-31 N,  54-55 E in the south-east of the city Yazd. At
18 Feb. 2004 at 5:00 GMT, we observed the second ionized currents
at the same location. At this location over the local granites
appearing out of the ground, we measured the surface resistivity
of the granites using the four point probe method. The existence
of positive potential on the granite surface was
 detected. Using the MODIS on board NASA's Terra satellite
 data we checked the mid-IR radiation of the location in channel
 20, with wavelength range of 3.660-3.840 $\mu m$.
We checked that from Nov. 2003 until 10th Feb. 2004, there was no
difference in the gray level distribution of the mid-IR
radiation. At 15th Feb. 2004 the statistical properties of the
radiation has changed, for instance the standard deviation of
intensity distribution changes from 24.41 at 10th Feb. to 11.47
at 15th Feb. The situation remained stable until to 24th Feb.
2004. At 24th Feb. the standard deviation was 16.58 and the
radiation location goes to turn off. At 25th Feb. 2004 the whole
statistical properties has returned to the situation similar to
Nov. 2003. At 26th Feb. 2004 at 20:10 GMT the first shocks of the
earthquake hit the the region. On 27th Feb. 2004 at 04:40 GMT and
04:45 GMT, the region has been hit by two other earthquakes with
3.1M. At 28th Feb. 2004 at 00:34 GMT, the region was hit by
another earthquake with magnitude 2.9M and then on 28th Feb. 2004
at 01:11 GMT by 3.7 M.

\begin{figure}
\epsfxsize=8truecm\epsfbox{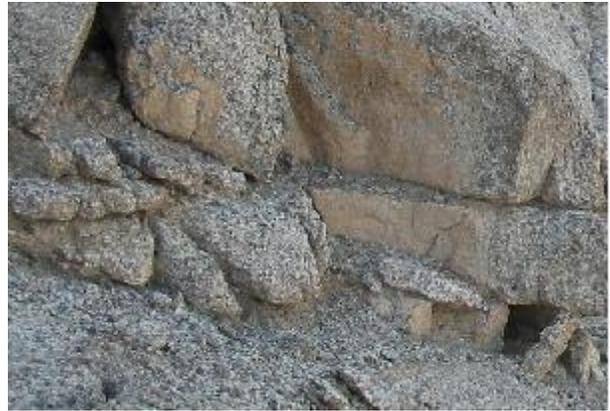} \narrowtext \caption{ The
granite appearing out of the ground close to Bahadoran fault
south east of Yazd. }
\end{figure}

We thank  F. Ardalan, H. Arfaei, R. Ashtiani, K. Esfarjani,  V.
Karimipour, and R. Mansouri, for their useful discussions.

\end{document}